\documentclass[12pt, draftcls, onecolumn]{IEEEtran}

\ifCLASSINFOpdf
  \usepackage[pdftex]{graphicx}
  \DeclareGraphicsExtensions{.pdf,.jpeg,.png}
\else
  \usepackage[dvips]{graphicx}
\fi
\usepackage{epstopdf}
\usepackage[cmex10]{amsmath}
\usepackage{amssymb}
\usepackage{wasysym}
\usepackage{algorithm}
\usepackage{algorithmic}
\usepackage{array}
\usepackage{cite}
\usepackage{color}
\usepackage{url}

\usepackage{epsfig,latexsym}
\usepackage{flushend}
\usepackage{cite}
\usepackage{verbatim}
\usepackage{amsopn}
\usepackage{booktabs}

\usepackage{stfloats}
\usepackage{hyperref}

\begin{document}
%
\title{Codeword Selection and Hybrid Precoding for Multiuser Millimeter Wave Massive MIMO Systems}


\author{Xuyao~Sun,~\IEEEmembership{Student~Member,~IEEE}, and Chenhao~Qi,~\IEEEmembership{Senior~Member,~IEEE}
\thanks{This work is supported in part by National Natural Science Foundation of China under Grant 61871119 and Natural Science Foundation of Jiangsu Province under Grant BK20161428. (\textit{Corresponding author: Chenhao~Qi})}
\thanks{Xuyao~Sun and Chenhao~Qi are with the School of Information Science and Engineering, Southeast University, Nanjing 210096, China (Email: qch@seu.edu.cn).}
}

\markboth{}
{Shell \MakeLowercase{\textit{et al.}}: Bare Demo of IEEEtran.cls for Journals}

\maketitle

\begin{abstract}
Aiming at maximizing the achievable sum-rate of wideband multiuser mmWave massive MIMO systems, the hybrid precoding is studied. Since each computation of the achievable sum-rate can be performed only after the analog precoder and digital precoder are both determined, the maximization of the achievable sum-rate has intractable computational complexity. By introducing the interference free (IF) achievable sum-rate, the design of the analog and digital precoders can be decoupled. To avoid the beam conflict and maximize the IF achievable sum-rate, a Hungarian-based codeword selection algorithm is proposed for the analog precoding design. Simulation results verify the effectiveness of the proposed scheme and show that better performance can be achieved compared with existing schemes.

\end{abstract}

\section{Introduction}
Millimeter wave (mmWave) massive multi-input multi-output (MIMO), which can achieve multi-gigabit data rates benefiting from its abundant frequency resource, has become an attractive candidate for future wireless communication techniques~\cite{Han2015}. However, recent work shows that the wideband mmWave massive MIMO channel is frequency selective~\cite{Heath2018}. To deal with the frequency selective channel, orthogonal frequency division multiplexing (OFDM) is typically adopted~\cite{Heath2018}. With OFDM, the frequency selective channel is transformed into several frequency flat subchannels, where each subchannel corresponds to an OFDM subcarrier.  Consequently, the challenging problem is the design of hybrid analog-digital precoders for the wideband channel, which is substantially different with that for the narrowband channel, since for the wideband channel the analog precoders should be jointly designed for all OFDM subcarriers while the digital precoders are designed individually.

Given the perfect channel state information (CSI) and fully digital precoding matrix, some work studies the design of hybrid precoding matrix to approach the fully digital precoding matrix. In~\cite{yu2016alternating}, the hybrid precoding design is treated as a matrix factorization problem, where alternating minimization algorithms are proposed for both the fully-connected and partially-connected structures. In~\cite{Ayach2014}, the hybrid precoding design is formulated as a sparse reconstruction problem, where an orthogonal matching pursuit (OMP) based spatially sparse precoding algorithm is proposed and can be further extended to be the simultaneous OMP (SOMP) based algorithm~\cite{Hou2014}. Aiming at maximizing the achievable sum-rate, a two-stage limited feedback multiuser hybrid precoding algorithm is proposed in~\cite{alkhateeb2015limited}. However, \cite{alkhateeb2015limited} does not consider multiuser beam conflict, where different users may be communicated by the base station (BS) with the same analog beam, leading to the low rank of the analog precoder matrix and consequently the reduction of the achievable sum-rate.


In this letter, aiming at maximizing the achievable sum-rate, we propose a hybrid precoding scheme for wideband multiuser mmWave massive MIMO system. Since each computation of the achievable sum-rate can be performed only after the analog precoder and digital precoder are both determined, the maximization of the achievable sum-rate has intractable computational complexity. By introducing the interference free (IF) achievable sum-rate, we decouple the design of the analog and digital precoders. To avoid the beam conflict and maximize the IF achievable sum-rate, we propose a Hungarian-based codeword selection algorithm for the analog precoding design. Note that the codeword selection and hybrid precoding for narrowband channel cannot be directly applied for wideband channel. To our best knowledge, this work is the first one to treat the beam conflict in wideband multiuser mmWave massive MIMO system.


\section{Problem Formulation}\label{sec.ProblemFormulation}
Consider a wideband multiuser mmWave massive MIMO system with a base station (BS) and $U$ users. The BS is equipped with $N$ uniform linear array (ULA) antennas and $S$ RF chains ($N\gg S\ge 1$), while each user is equipped with a single antenna. Based on orthogonal multiple access which is commonly used in the existing multiuser wireless systems, the maximum number of users simultaneously served by the BS is restricted by the number of its RF chains, i.e., $U\leq S$. We assume that the BS and users are fully synchronized and time division duplex (TDD) is adopted in this work.

To deal with the frequency selective fading, the mmWave massive MIMO system normally uses OFDM~\cite{Heath2018}. Suppose the number of OFDM subcarriers is $K$. The BS adopts a hybrid precoder including an analog precoder $\boldsymbol{F}_{\rm RF}$ and $K$ digital precoders $\boldsymbol{F}_{\rm BB}[k],k=0,...,K-1$. Note that the analog precoder is the same for different subcarriers, since the angle of departure (AoD) and angle of arrival (AoA) for different subcarriers are the same and therefore are independent of the frequency. The system model for downlink transmission using the $k$th OFDM subcarrier can be expressed as
\begin{equation}\label{finaldata}
\boldsymbol{y}[k]=\boldsymbol{H}[k]\boldsymbol{F}_{\rm RF}\boldsymbol{F}_{\rm BB}[k]\boldsymbol{s}[k]+\boldsymbol{z}[k
],
\end{equation}
where $\boldsymbol{s}[k]\in\mathbb{C}^{U\times 1}$ is the transmitted signal using the $k$th OFDM subcarrier with total power $\rho$, i.e., $\mathbb{E}\big\{\boldsymbol{s}[k]\boldsymbol{s} ^{H}[k]\big\}=\frac{\rho}{KU}\boldsymbol{I}_{U}$. The received signal by the $k$th OFDM subcarrier is denoted as $\boldsymbol{y}[k]\in\mathbb{C}^{U\times 1}$, where the $u(u\in \boldsymbol{\mathcal{U}}\triangleq\{1,2,\ldots,U\})$th entry of $\boldsymbol{y}[k]$ is the signal received by the $u$th user. The channel matrix is denoted as $\boldsymbol{H}^T[k]  \triangleq {\big[\boldsymbol{h}^T_1[k],\ldots,\boldsymbol{h}^T_U[k]\big]}$, where $\boldsymbol{h}_{u}[k] \in \mathbb{C}^{1\times N}$ is the channel vector between the BS and the $u$th user. $\boldsymbol{z}[k]\sim\mathcal{CN}(0,\sigma^2\boldsymbol{I}_U)$ denotes the additive white Gaussian noise with zero mean and covariance being $\sigma^2$.
$\boldsymbol{F}_{\rm BB}[k] \triangleq \big[\boldsymbol{f}_{\rm BB}^{1}[k],\boldsymbol{f}_{\rm BB}^{2}[k],\ldots,\boldsymbol{f}_{\rm BB}^{U}[k]\big] \in \mathbb{C}^{S \times U } $ denotes the digital precoder, where $\boldsymbol{f}_{\rm BB}^{u}[k]$ denotes the $u$th column of $\boldsymbol{F}_{\rm BB}[k]$. $\boldsymbol{F}_{\rm RF} \triangleq [\boldsymbol{f}_{\rm RF}^{1},\boldsymbol{f}_{\rm RF}^{2},\ldots,\boldsymbol{f}_{\rm RF}^{S}] \in \mathbb{C}^{N\times S} $ denotes the analog precoder achieved by a phase-shifter network, where each entry of $\boldsymbol{F}_{\rm RF}$ has a constant modulus and only changes its phase. Moreover, regarding that the hybrid precoder does not provide power gain, we set ${\|\boldsymbol{F}_{\rm RF}\boldsymbol{F}_{\rm BB}[k]\|}^2_F=S$ to satisfy the power normalization. For simplicity, we assume $S=U$ in the rest of this letter.

The channel between the BS and each user is assumed to be frequency-selective, having a maximum delay of $N_c$ taps in the time domain. We adopt the geometric wideband mmWave MIMO channel model~\cite{Heath2018}, where the channel between the BS and the $u$th user with a delay of $d(d=0,1,\ldots,N_c-1)$ taps can be expressed as
\begin{small}
\begin{equation}\label{channelmodel}
\boldsymbol{h}_u^{(d)}=\sqrt{\frac{N}{L\beta_{u}}}\sum_{l=1}^{L}\alpha_{l,u}p(dT_s-\tau_{l,u})\boldsymbol{a}^H(N,\theta_{l,u}),
\end{equation}
\end{small}where $\beta_{u}$, $L$, $\alpha_{l,u}$, $\tau_{l,u}$, $p(\tau)$ and $T_s$ denotes the path loss, the number of paths, the complex gain for the $l$th path, the delay of the $l$th path, the pulse-shaping filter observed at $\tau$ and the sampling period, respectively. $\theta_{l,u}\in[-\pi/2,\pi/2)$ denotes the AoD of the $l$th path. The channel steering vector $\boldsymbol{a}(N,\theta)\in \mathbb{C}^{N\times1}$ is expressed as
\begin{small}
\begin{equation}\label{BSchannelresponsevector}
\boldsymbol{a}(N,\theta)=\frac{1}{\sqrt{N}}{\Big[1,e^{j\frac{2\pi m}{\lambda}\sin(\theta)},\ldots,e^{j\frac{2\pi m}{\lambda}(N-1)\sin(\theta)}\Big]}^T
\end{equation}
\end{small}where $\lambda$ is the signal wavelength and $m$ is the antenna spacing. Based on~(\ref{channelmodel}), $\boldsymbol{h}_{u}[k]$ can be written as
\begin{small}
\begin{equation}\label{channelmodelk}
\boldsymbol{h}_{u}[k]=\sum_{d=0}^{N_c-1}\boldsymbol{h}_u^{(d)}e^{-j\frac{2\pi k}{K}d}
\end{equation}
\end{small}which is essentially the $K$-point discrete Fourier transform (DFT).

Based on (\ref{finaldata}), the achievable rate for the $u$th user on the $k$th subcarrier is expressed as
\begin{small}
\begin{equation}\label{AchievableRate}
R_u[k]=\log_{2}\bigg(1+\frac{\frac{\rho}{KU}|\boldsymbol{h}_u[k]\boldsymbol{F}_{\rm RF}\boldsymbol{f}_{\rm BB}^u[k]|^2}{\frac{\rho}{KU}\sum_{i\neq U}|\boldsymbol{h}_u[k]\boldsymbol{F}_{\rm RF}\boldsymbol{f}_{\rm BB}^i[k]|^2+\sigma^2}\bigg).
\end{equation}
\end{small}This work aims at maximizing the achievable sum-rate of all the $U$ users, which can be formulated as
\begin{subequations}\label{ProblemFormulation}
\begin{align} \label{target}
&\max_{\boldsymbol{F}_{\rm RF}, \{\boldsymbol{F}_{\rm BB}[k]\}_{k=0}^{K-1}} \frac{1}{K}\sum^U_{u=1}\sum^{K-1}_{k=0}R_u[k]\\ \label{constraint1}
&{\rm s.t.}~\boldsymbol{f}^u_{\rm RF}\in\boldsymbol{\mathcal{F}}_c,\forall u\in \boldsymbol{\mathcal{U}},\\ \label{constraint2}
&~~~~~\boldsymbol{f}^i_{\rm RF}\neq\boldsymbol{f}^q_{\rm RF},\forall i,q\in \boldsymbol{\mathcal{U}},i\neq q,\\
&~~~~~{\|\boldsymbol{F}_{\rm RF}\boldsymbol{F}_{\rm BB}[k]\|}^2_F=U,k=0,...,K-1. \label{constraint3}
\end{align}
\end{subequations}
In this problem, (\ref{constraint1}) indicates that each column vector of $\boldsymbol{F}_{\rm RF}$ is selected from a predefined codebook $\boldsymbol{\mathcal{F}}_c$. The widely used DFT codebook is adopted, where $\boldsymbol{\mathcal{F}}_c\triangleq \{\boldsymbol{f}_c(n),\forall n\in \boldsymbol{\mathcal{N}}\triangleq\{1,2,\ldots,N\}\}$, with $\boldsymbol{f}_c(n) \triangleq \boldsymbol{a}(N,\sin^{-1}[-1+(2n-1)/N])$~\cite{Suh2017}. It is essentially to select $U$ out of $N$ codewords. Since each codeword can form a beam with the width of $2/N$, (\ref{constraint2}) makes sure that different codewords from $\boldsymbol{\mathcal{F}}_c$ are selected to serve different users, which can avoid the beam conflict. (\ref{constraint3}) indicates the power normalization for the hybrid precoder.


The optimization problem (\ref{ProblemFormulation}) is challenging due to the difficulty in multiuser codeword selection and the coupling between $\boldsymbol{F}_{\rm RF}$ and $\{\boldsymbol{F}_{\rm BB}[k]\}_{k=0}^{K-1}$. A straightforward method to select $U$ out of $N$ codewords is to test all the ${\tiny \begin{pmatrix}N\\U\end{pmatrix}}$ possibilities, which is prohibitively complex even for moderate $N$ and $U$. Furthermore, each test to compute the achievable sum-rate can be performed only after $\boldsymbol{F}_{\rm RF}$ and $\{\boldsymbol{F}_{\rm BB}[k]\}_{k=0}^{K-1}$ are both determined. Therefore, the maximization of the achievable sum-rate has intractable computational complexity.

\section{Hybrid Precoding}\label{sec.ProsedHybridPrecodingDesign}
In this section, by introducing the interference free (IF) achievable rate, we decouple the design of the analog precoder and digital precoder, which is involved in the sum-rate maximization problem expressed in \eqref{ProblemFormulation}. We propose a hybrid precoding scheme, where a Hungarian-based codeword selection and analog precoding algorithm is proposed.


It is seen that there is multiuser interference in the denominator of \eqref{AchievableRate}. Since \eqref{constraint2} guarantees there is no beam conflict, the equivalent channel for the $k$th OFDM subcarrier, denoted by $\boldsymbol{H}[k]\boldsymbol{F}_{\rm RF}$ is full rank. Then the multiuser interference can be completely removed after using the digital precoder $\boldsymbol{F}_{\rm BB}[k]$. Therefore, in the context of no multiuser interference, the digital precoder can be temporarily neglected, leading the achievable rate for the $u$th user on the $k$th subcarrier to be
\begin{small}
\begin{equation}\label{IFAchievableRate}
R_{u}^{\rm IF}[k]=\log_{2}\Bigg(1+\frac{\rho|\boldsymbol{h}_u[k]\boldsymbol{f}_{\rm RF}^u|^2}{KU\sigma^2}\Bigg).
\end{equation}
\end{small}where the superscript ``IF'' is short for interference free. In fact, \eqref{IFAchievableRate} represents the achievable rate for the single user scenario. Maximizing \eqref{IFAchievableRate} is consistent with the work maximizing the equivalent channel gain~\cite{alkhateeb2015limited}. However, \cite{alkhateeb2015limited} did not consider the beam conflict. In this way, the analog precoder design that is mainly determined by the codeword selection to avoid beam conflict, is decoupled from the digital precoder design. Now we resort to the following problem expressed as
\begin{subequations}\label{Subproblem1}
\begin{align}
&\max_{\boldsymbol{F}_{\rm RF}} \frac{1}{K} \sum^U_{u=1}\sum^{K-1}_{k=0}R^{\rm IF}_u[k] \label{ObjSummationMax}\\
&~{\rm s.t.}~\boldsymbol{f}^u_{\rm RF}\in\boldsymbol{\mathcal{F}}_c,\forall u\in \boldsymbol{\mathcal{U}}, \\
&~~~~~~\boldsymbol{f}^i_{\rm RF}\neq\boldsymbol{f}^q_{\rm RF},\forall i,q\in \boldsymbol{\mathcal{U}},i\neq q.
\end{align}
\end{subequations}
It is a typical codeword selection problem to maximize the IF achievable sum-rate. To solve this problem, we propose a
Hungarian-based codeword selection algorithm, which is summarized in~\textbf{Algorithm~\ref{alg1}}.

First we define a matrix $\boldsymbol{T}\in\mathbb{C}^{U\times N}$, where the entry on the $u$th row and $n$th column of $\boldsymbol{T}$, denoted as $\boldsymbol{T}(u,n)$, $u\in \boldsymbol{\mathcal U}$, $n\in \boldsymbol{\mathcal N}$, is expressed as
\begin{small}
\begin{equation}\label{CostMatrix}
\boldsymbol{T}(u,n)\triangleq \mathcal{I}\Bigg(
\frac{1}{K}\sum_{k=0}^{K-1}\log_{2}\Big(1+\frac{\rho|\boldsymbol{h}_u[k]\boldsymbol{f}_c(n)|^2}{KU\sigma^2}\Big),\gamma_u\Bigg)
\end{equation}
\end{small}where the function $\mathcal{I}(x,y)$ with two variables $x$ and $y$ is defined as
\begin{equation}\label{FunctionDefine}
    \mathcal{I}(x,y)\triangleq \left\{ \begin{array}{cl}
	x, &x \geq y,\\
	0,& \textrm{else}.
\end{array} \right.
\end{equation}
$\gamma_u$ is a threshold to control the number of candidate codewords for the $u$th user. In fact, $\boldsymbol{T}(u,n)$ is the IF achievable rate of the $u$th user with the codeword $\boldsymbol{f}_c(n)$ if it is larger than $\gamma_u$. Otherwise, $\boldsymbol{T}(u,n)$ is zero indicating $\boldsymbol{f}_c(n)$ is not a candidate codeword for the $u$th user. During the beam training, we set each of $\boldsymbol{f}_c(n),n\in \boldsymbol{\mathcal N}$ as the analog precoding vector to serve the $u$th user, obtaining $N$ different achievable rate. Given the number of candidate codewords for the $u$th user, denoted as $M_u(M_u \geq 2)$, we set $\gamma_u$ as the $M_u$th largest value of all the obtained $N$ achievable sum-rate. The motivation of using more than one candidate codeword for each user is to provide at least one more option for each user when the beam conflict happens, i.e., different users select the same codeword as their analog precoding vectors. On the other side, larger $M_u$ will lead to higher computational complexity.

\begin{algorithm}[!t]
	\caption{Hungarian-based Codeword Selection and Analog Precoding}
	\label{alg1}
	\begin{algorithmic}[1]
        \STATE \textbf{Input:} $N$, $U$, $\boldsymbol{\mathcal{F}}_c$, $\boldsymbol{T}$.
        \STATE Remove zero columns from $\boldsymbol{T}$, obtaining $\boldsymbol{T'}\in\mathbb{C}^{U\times N'}$.
        \STATE Obtain $t$ via \eqref{maxTmatrix}.
        \STATE Update $\boldsymbol{T'}$ via \eqref{updateTmatrix} and \eqref{ConvertDimension}.
        \STATE Use the Hungarian algorithm to solve \eqref{AssignmentProblem2} to obtain $\boldsymbol{P'}$.
        \STATE Update $\boldsymbol{P'}$ by removing the bottom $\Delta_{N}$ rows from $\boldsymbol{P'}$.
        \STATE Compute $\boldsymbol{P}$ corresponding to $\boldsymbol{P'}$.
        \STATE Select codewords from $\boldsymbol{\mathcal{F}}_c$ based on $\boldsymbol{P}$.
        \STATE Determine $\widetilde{\boldsymbol{F}}_{\rm RF}$ based on the selected codewords.
        \STATE \textbf{Output:} $\widetilde{\boldsymbol{F}}_{\rm RF}$.
	\end{algorithmic}
\end{algorithm}

We introduce a binary matrix $\boldsymbol{P}$, which is of the same dimension as $\boldsymbol{T}$. The indices of the finally selected codewords are stored in $\boldsymbol{P}$. If $\boldsymbol{f}_c(n)$ is finally selected to serve the $u$th user, then $\boldsymbol{P}(u,n)=1$; otherwise, $\boldsymbol{P}(u,n)=0$.

Then \eqref{Subproblem1} can be rewritten as
\begin{small}
\begin{subequations}\label{AssignmentProblem}
\begin{align}
&\max_{\boldsymbol{P}}~\sum^U_{u=1}\sum^N_{n=1}\boldsymbol{T}(u,n) \boldsymbol{P}(u,n),\label{maxTandP} \\
&~{\rm s.t.}\sum^U_{u=1}\boldsymbol{P}(u,n)\leq1, n\in \boldsymbol{\mathcal{N}};~\sum^N_{n=1}\boldsymbol{P}(u,n)=1, u\in \boldsymbol{\mathcal{U}}\label{ConstraintForEachUser}.
\end{align}
\end{subequations}
\end{small}To avoid the beam conflict, \eqref{ConstraintForEachUser} indicates that each codeword can be selected at most once and each user can select only one codeword.

Before applying the original Hungarian algorithm~\cite{Wang2011}, we make some preprocessing on $\boldsymbol{T}$. As shown in the second step of \textbf{Algorithm~\ref{alg1}}, we remove zero columns from $\boldsymbol{T}$, resulting in $\boldsymbol{T'}$. The purpose of this step is to reduce the storage overhead as well as the computational complexity thereafter. Suppose the number of columns of $\boldsymbol{T'}$ is $N'(U\leq N'\leq N)$. Note that $U\leq N'$ should be satisfied so that the number of all candidate codewords is greater than the number of users. If $U\leq N'$ can not be satisfied, we should remove less zero columns from $\boldsymbol{T}$. We define $\boldsymbol{\mathcal{N}'}\triangleq\{1,2,...,N'\}$. As shown in the third step of \textbf{Algorithm~\ref{alg1}}, we obtain the largest entry from $\boldsymbol{T'}$ as
\begin{equation}\label{maxTmatrix}
  t= \max_{u \in \boldsymbol{\mathcal{U}}, n \in \boldsymbol{\mathcal{N}'}} \boldsymbol{T'}(u,n)
\end{equation}
Then we update $\boldsymbol{T'}$ by
\begin{equation}\label{updateTmatrix}
\boldsymbol{T'}(u,n) \leftarrow t-\boldsymbol{T'}(u,n),~\forall u \in \boldsymbol{\mathcal{U}}, \forall n \in \boldsymbol{\mathcal{N}'}.
\end{equation}
Note that the original Hungarian algorithm can only deal with the summation minimization problem~\cite{Wang2011}. Therefore we use \eqref{updateTmatrix} to convert the summation maximization problem expressed in \eqref{maxTandP} to a summation minimization problem. Since the Hungarian algorithm can only tackle the square matrix, we use \eqref{ConvertDimension} to convert $\boldsymbol{T'}$ into a square matrix as
\begin{equation}\label{ConvertDimension}
 \boldsymbol{T'} \leftarrow [\boldsymbol{T'}; \boldsymbol{0}^{\Delta_{N}\times N'} ]
\end{equation}
which is essentially to add $\Delta_{N}(\Delta_{N} \triangleq N'-U)$ zero rows at the bottom of $\boldsymbol{T'}$. Once finishing the preprocessing from step~2 to step~4, we can convert \eqref{AssignmentProblem} into the following problem.
\begin{footnotesize}
\begin{subequations}\label{AssignmentProblem2}
\begin{align}
&\min_{\boldsymbol{P'}}~\sum^{N'}_{u=1}\sum^{N'}_{n=1}\boldsymbol{T'}(u,n) \boldsymbol{P'}(u,n),\label{maxTandP2} \\
&~{\rm s.t.}\sum^{N'}_{u=1}\boldsymbol{P'}(u,n)=1, n\in \boldsymbol{\mathcal{N'}};~\sum^{N'}_{n=1}\boldsymbol{P'}(u,n)=1, u\in \boldsymbol{\mathcal{N'}}\label{ConstraintForEachUser2}.
\end{align}
\end{subequations}
\end{footnotesize}where $\boldsymbol{P'}$ is a binary matrix in the same dimension as $\boldsymbol{T'}$. Note that~\eqref{AssignmentProblem2} is a typical assignment problem, the original Hungarian algorithm can be used to obtain its optimal solution, where $\boldsymbol{T'}$ and $\boldsymbol{P'}$ are the input and output of the Hungarian algorithm, respectively.

We briefly describe the main steps of the Hungarian algorithm.
1) Given $\boldsymbol{T'}$, we subtract the entries of each row by the smallest entry of that row, to produce at least one zero entry on each row. Similarly, we subtract the entries of each column by the smallest entry of that column, to produce at least one zero entry on each column.
2) We use rows and columns to cover all the zero entries aiming at minimizing the summation of the number of rows and columns until the summation equals $N'$. If the summation is smaller than $N'$, we iteratively perform the following procedures. 2.1) We find the smallest entry among all the uncovered entries. 2.2) We subtract the entries in all uncovered rows by this smallest entry. We add the entries in all covered columns by this smallest entry. 2.3) We use the smallest summation number of rows and columns to cover all the zero entries.
3) We output a full-rank binary matrix $\boldsymbol{P'}$ including $N'$ nonzero entries in total, where each nonzero entry of $\boldsymbol{P'}$ corresponds to a zero entry of $\boldsymbol{T'}$.

After running the Hungarian algorithm, we update $\boldsymbol{P'}$ by removing the bottom $\Delta_{N}$ rows from $\boldsymbol{P'}$, since these $\Delta_{N}$ rows correspond to the $\Delta_{N}$ zero rows in \eqref{ConvertDimension}. In this way, the number of nonzero entries in $\boldsymbol{P}$ is $U$. Then we compute $\boldsymbol{P}$ corresponding to $\boldsymbol{P'}$, which is essentially to find the coordinates of nonzero entries of $\boldsymbol{P'}$ and set them in the corresponding coordinates of $\boldsymbol{P}$. For each nonzero entry of $\boldsymbol{P}$, the column index and the row index are denoted as $\tilde{n}$ and $\tilde{u}$, respectively. Then the selected codeword from $\boldsymbol{\mathcal{F}}_c$ for the $\tilde{u}$th user is $\footnotesize\widetilde{\boldsymbol{f}}^{\tilde{u}}_{\rm RF}=\boldsymbol{f}_c(\tilde{n})$. Finally, the designed analog precoder is $\footnotesize \widetilde{\boldsymbol{F}}_{\rm RF}=[\widetilde{\boldsymbol{f}}^1_{\rm RF},\widetilde{\boldsymbol{f}}^2_{\rm RF},\ldots,\widetilde{\boldsymbol{f}}^U_{\rm RF}]$.

Given $\footnotesize\widetilde{\boldsymbol{F}}_{\rm RF}$, for the $k$th subcarrier, we define the equivalent channel matrix $\footnotesize\boldsymbol{H}_{\rm e}[k]=\boldsymbol{H}[k]\widetilde{\boldsymbol{F}}_{\rm RF}$ which is essentially the multiplication of the channel matrix and the analog precoder. To estimate $\footnotesize\boldsymbol{H}_{\rm e}[k]$, pilot-assisted channel estimation is typically performed~\cite{Zhao2017}. Suppose an estimate of $\footnotesize\boldsymbol{H}_{\rm e}[k]$ is denoted as $\footnotesize\widetilde{\boldsymbol{H}}_{\rm e}[k]$. Then the digital precoders based on the zero-forcing (ZF) criterion or minimum mean square error (MMSE) criterion can be determined by $\footnotesize\boldsymbol{F}^{\rm ZF}_{\rm BB}[k]={\widetilde{\boldsymbol{H}}}_{\rm e}^H[k]\big({\widetilde{\boldsymbol{H}}}_{\rm e}[k]{\widetilde{\boldsymbol{H}}}^H_{\rm e}[k]\big)^{-1}$ and $\footnotesize\boldsymbol{F}^{\rm MMSE}_{\rm BB}[k]={\widehat{\boldsymbol{H}}}_{\rm e}^H[k] \big[(\rho/KU){\widetilde{\boldsymbol{H}}}_{\rm e}[k]{\widetilde{\boldsymbol{H}}}_{\rm e}^H[k]+\sigma^2\boldsymbol{I}_{U}\big]^{-1}$, respectively. Finally, we normalize each column of $\footnotesize\boldsymbol{F}^{\rm ZF}_{\rm BB}[k]$ or $\footnotesize\boldsymbol{F}^{\rm MMSE}_{\rm BB}[k], k=0,\ldots,K-1$, to satisfy the power normalization in \eqref{constraint3}.

\section{Simulation Results}\label{sec.SimulationResults}
\begin{figure}[!t]
\centering
\includegraphics[height=68mm]{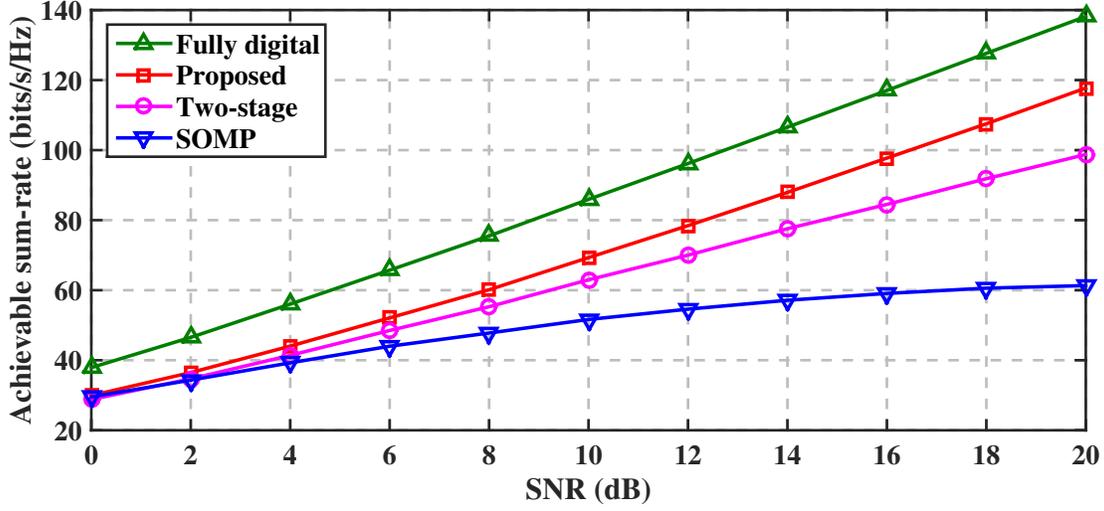}
\caption{Comparison of achievable sum-rate for different precoding schemes in terms of different SNR.}
\label{fig:SNR}
\end{figure}
\begin{figure}[!t]
\centering
\includegraphics[height=68mm]{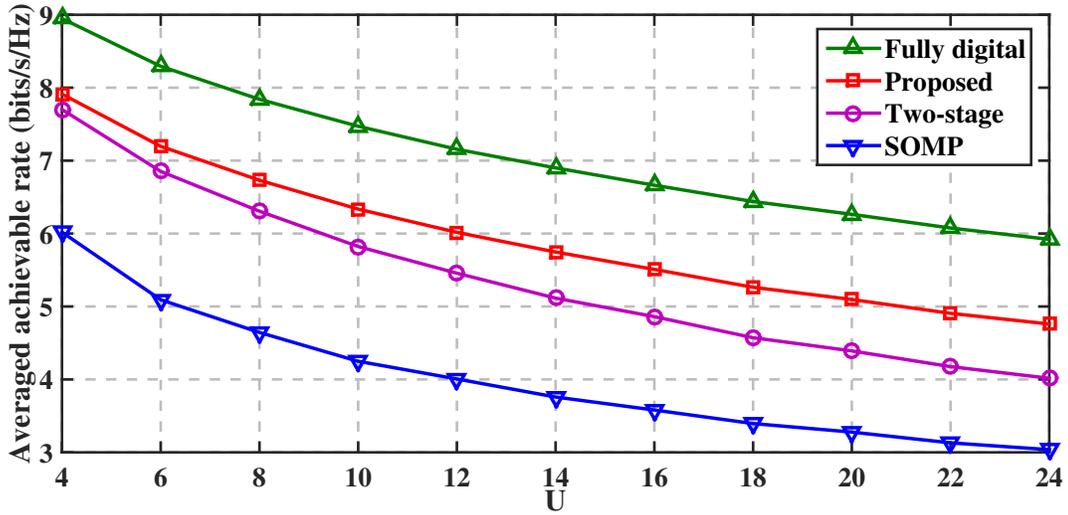}
\caption{Comparison of averaged achievable rate for different precoding schemes in terms of different $U$.}
\label{fig:K}
\end{figure}
\begin{table}[!t]
\centering
\caption{Comparisons of running time for different precoding schemes in terms of different $U$ (in $10^{-3}$ Second).}
\label{Table1}
\begin{tabular}{p{2.1cm}p{1.5cm}p{1.5cm}p{1.5cm}p{1.5cm}p{1.5cm}p{1.5cm}p{1.5cm}}
\toprule
 &   $U=4$     &   $U=8$      & $U=12$ &   $U=16$     &   $U=20$     &   $U=24$    \\
\midrule
Fully digital  &   2.25  &   2.43  &   3.45  &   4.26  &   5.27  &   5.98  \\
Two-stage &   3.16  &   5.91  &   8.80  &   11.75  &   14.81  &   17.98  \\
SOMP  &   21.83  &   40.12  &   72.93  &   109.10  &   167.10  &   213.54  \\
Proposed &   3.29  &   6.63  &   12.62  &  22.16  &   35.18  &   53.10  \\
\bottomrule
\end{tabular}
\end{table}

The considered wideband multiuser mmWave massive MIMO system includes a BS and $U=16$ single-antenna users. The BS is equipped with $N=128$ ULA antennas and $S=U=16$ RF chains. The number of OFDM subcarriers is $K=16$. The wideband channels are generated according to the $N_c$ delay model in~\cite{Heath2018} with $N_c=4$, $L=4$, $\beta_u=0.25$, and $T_s=\frac{1}{1760}\mu s$. We set $\alpha_{1,u}\sim\mathcal{CN}(0,1)$ and $\alpha_{l,u}\sim\mathcal{CN}(0,0.1)$ for $l=2,3,4$. Suppose $\tau_{l,u}$ obeys the uniform distribution in $[0,(N_c-1)T_s]$ and $\theta_{l,u}$ obeys the uniform distribution in $[-\pi/2,\pi/2)$. For each user, we select $M_u=4$ candidate codewords for the proposed codeword selection algorithm. The MMSE criterion is adopted for the digital precoder design. We run the Monte Carlo simulations with 2000 random channel implementations.

As shown in Fig.~(\ref{fig:SNR}), we compare the achievable sum-rate for different precoding schemes in terms of different SNR. The comparison includes the proposed scheme, the existing SOMP scheme~\cite{Ayach2014,Hou2014} and two-stage scheme~\cite{alkhateeb2015limited}. We set ${\rm SNR}=\rho / (U\sigma^2)$. The performance of fully digital precoding is also provided as the upper bound. It is seen that the proposed scheme outperforms the SOMP scheme and two-stage scheme, since SOMP scheme does not completely remove the multiuser interference and the two-stage scheme does not consider the beam conflict. In particular, at ${\rm SNR}=14$ dB, the proposed scheme has 13.38\% and 56.92\% improvement over the two-stage scheme and the SOMP scheme, respectively.


As shown in Fig.~(\ref{fig:K}), we compare the averaged achievable rate for different precoding schemes in terms of different $U$. The averaged achievable rate is defined as the ratio of the achievable sum-rate over the number of users $U$. We fix downlink ${\rm SNR}=14$ dB. It is seen that as $K$ increases, the proposed scheme drops more slowly than the SOMP scheme and the two-stage scheme.

We also compare the running time for different precoding schemes in Table~\ref{Table1}. The simulations are performed using MATLAB R2014b running on a desktop with four Intel Core i5-3470 CPUs at 3.6 GHz and 8GB memory.
The running time for different precoding schemes get longer as $U$ increases. It is seen that the fully digital precoding is the fastest among the four schemes, but it can not be implemented in practice.  The proposed scheme is faster than the SOMP scheme but is slower than the two-stage scheme, since the codeword selection occupying additional computational resource is included in the proposed scheme but not included in the two-stage scheme.

\section{Conclusions}\label{sec.conclusion}
We have proposed a hybrid precoding scheme aiming at maximizing the sum-rate of wideband multiuser mmWave massive MIMO systems. To avoid the beam conflict, a Hungarian-based codeword selection and analog precoding algorithm has been proposed. Future work will be continued with the focus on the scenario with more users than the BS RF chains.

\bibliographystyle{IEEEtran}
\bibliography{IEEEabrv,IEEEexample}

\begin{thebibliography}{1}
\providecommand{\url}[1]{#1}
\csname url@samestyle\endcsname
\providecommand{\newblock}{\relax}
\providecommand{\bibinfo}[2]{#2}
\providecommand{\BIBentrySTDinterwordspacing}{\spaceskip=0pt\relax}
\providecommand{\BIBentryALTinterwordstretchfactor}{4}
\providecommand{\BIBentryALTinterwordspacing}{\spaceskip=\fontdimen2\font plus
\BIBentryALTinterwordstretchfactor\fontdimen3\font minus
  \fontdimen4\font\relax}
\providecommand{\BIBforeignlanguage}[2]{{%
\expandafter\ifx\csname l@#1\endcsname\relax
\typeout{** WARNING: IEEEtran.bst: No hyphenation pattern has been}%
\typeout{** loaded for the language `#1'. Using the pattern for}%
\typeout{** the default language instead.}%
\else
\language=\csname l@#1\endcsname
\fi
#2}}
\providecommand{\BIBdecl}{\relax}
\BIBdecl

\bibitem{Han2015}
S.~Han, C.~I, Z.~Xu, and C.~Rowell, ``Large-scale antenna systems with hybrid
  analog and digital beamforming for millimeter wave {5G},'' \emph{{IEEE}
  Commun. Mag.}, vol.~53, no.~1, pp. 186--194, Jan. 2015.

\bibitem{Heath2018}
J.~Rodr¨ªguez-Fern¨¢ndez, N.~Gonz¨¢lez-Prelcic, K.~Venugopal, and R.~W. Heath,
  ``Frequency-domain compressive channel estimation for frequency-selective
  hybrid millimeter wave {MIMO} systems,'' \emph{{IEEE} Trans. Wireless
  Commun.}, vol.~17, no.~5, pp. 2946--2960, May 2018.

\bibitem{yu2016alternating}
X.~Yu, J.-C. Shen, J.~Zhang, and K.~B. Letaief, ``{Alternating minimization
  algorithms for hybrid precoding in millimeter wave MIMO systems},''
  \emph{IEEE J. Sel. Top. Signal Process.}, vol.~10, no.~3, pp. 485--500, Apr.
  2016.

\bibitem{Ayach2014}
O.~E. Ayach, S.~Rajagopal, S.~Abu-Surra, Z.~Pi, and R.~W. Heath, ``Spatially
  sparse precoding in millimeter wave {MIMO} systems,'' \emph{{IEEE} Trans.
  Wireless Commun.}, vol.~13, no.~3, pp. 1499--1513, Mar. 2014.

\bibitem{Hou2014}
W.~Hou and C.~W. Lim, ``Structured compressive channel estimation for
  large-scale {MISO-OFDM} systems,'' \emph{{IEEE} Commun. Lett.}, vol.~18,
  no.~5, pp. 765--768, May 2014.

\bibitem{alkhateeb2015limited}
A.~Alkhateeb, G.~Leus, and R.~W. Heath, ``Limited feedback hybrid precoding for
  multi-user millimeter wave systems,'' \emph{{IEEE} Trans. Wireless Commun.},
  vol.~14, no.~11, pp. 6481--6494, July 2015.

\bibitem{Suh2017}
J.~Suh, C.~Kim, W.~Sung, J.~So, and S.~W. Heo, ``Construction of a generalized
  {DFT} codebook using channel-adaptive parameters,'' \emph{{IEEE} Commun.
  Lett.}, vol.~21, no.~1, pp. 196--199, Jan. 2017.

\bibitem{Wang2011}
Z.~Wang, Z.~Feng, and P.~Zhang, ``An iterative {Hungarian} algorithm based
  coordinated spectrum sensing strategy,'' \emph{{IEEE} Commun. Lett.},
  vol.~15, no.~1, pp. 49--51, Jan. 2011.

\bibitem{Zhao2017}
L.~Zhao, D.~W.~K. Ng, and J.~Yuan, ``Multiuser precoding and channel estimation
  for hybrid millimeter wave systems,'' \emph{IEEE J. Sel. Areas Commun.},
  vol.~35, no.~7, pp. 1576--1590, Jul. 2017.

\end{thebibliography}

\end{document}